\newcommand{\gd}{\dot{\gamma}} 
\newcommand{\mg}{\overline{\gd}}  
\newcommand{\hop}{\omega}      

\documentclass[preprint]{epl}

\usepackage{psfig}

\title{Rheological instability in a simple shear thickening model}

\shorttitle{Rheological instability in a simple model}

\author{D. A. Head\inst{1} \and A. Ajdari\inst{2} \and M. E. Cates\inst{1}}
\institute{
\inst{1}Department of Physics and Astronomy, JCMB King's Buildings,
University of Edinburgh, Edinburgh EH9 3JZ, UK\\
\inst{2}Laboratoire de Physico--Chimie Th\'eorique, Esa CNRS 7083,
ESPCI, 10 rue Vauquelin, F-75231 Paris Cedex 05, France
}
\pacs{83.60.Rs}{Shear rate-dependent structure (shear thinning and shear thickening)}
\pacs{83.60.Wc}{Flow instabilities}
\pacs{82.40.Bj}{Oscillations, chaos, and bifurcations}

\begin{document}

\maketitle

\begin{abstract}
We study the strain response to steady imposed stress in a
spatially homogeneous,
scalar model for shear thickening, in which the local rate of
yielding $\Gamma(l)$ of mesoscopic `elastic elements' is not monotonic
in the local strain $l$. Despite this, the macroscopic, steady-state
flow curve (stress vs. strain rate) is monotonic. However, for a broad
class of $\Gamma(l)$, the response to steady stress is not in fact steady
flow, but spontaneous oscillation. We discuss this finding in relation to
other theoretical and experimental flow instabilities. Within the parameter
ranges we studied, the model does not exhibit rheo-chaos. 
\end{abstract}


The flow behaviour of shear-thickening materials such as dense colloidal
suspensions can be complex \cite{Laun,Melrose}. For example, imposition
of a steady mean strain rate can lead to large, possibly chaotic, variations
in the mean stress \cite{Laun}. The same occurs in some types of
shear-thickening micellar surfactant solutions, where true temporal chaos
seems now to be established \cite{Sood2}
(and also in shear thinning systems; see \cite{Sood1}).
Other unexpected behaviour,
such as a bifurcation to an oscillatory state, has also been seen in
shear-thickening `onion' phases of surfactant \cite{Jacques}.
It is not yet known to what extent such unsteady flow is generic in
shear-thickening systems; in this letter we attempt to shed some light
on the issue by studying a much-simplified, generic model. In this model
we find, for a wide range of parameters, spontaneous rheological
oscillation of the strain rate at fixed stress. Rheo-chaos is,
however, not found for the parameters studied so far. 

A feature that distinguishes the rheological instabilities encountered
in shear-thickening from those arising in Newtonian fluids is that the
nonlinearity is not inertial (not from the advective term of the Navier
Stokes equation): the Reynolds number is essentially zero~\cite{Groisman}.
Instead it
arises from anharmonic elastic responses at large deformations, complicated
and perhaps strongly enhanced by the presence of memory effects.
Flow instabilities leading to chaos have been studied recently by
Grosso {\em et al.}~\cite{Keunings}
in a model for suspended rodlike particles.
As that work shows, and our work confirms, temporal
instabilities can arise even in a model where macroscopic spatial
inhomogeneity is disallowed altogether. This is a strong demarcation
from the familiar shear-banding instabilities that arise whenever the
steady-state flow curve is nonmonotonic (the flow curve is
the function $\sigma(\dot\gamma)$,
where $\sigma$ is shear stress and $\dot\gamma$
the rate of shear strain) \cite{Ball,Spenley}. Shear banding is well
documented in shear thinning materials (particularly micelles
\cite{Spenley}) but also possible in shear thickening ones
\cite{Olmsted}, albeit with a different spatial organization
of the bands of coexisting material. However it is not the
subject of this Letter -- we are interested in {\em temporal} inhomogeneity.

Our simplified model of a shear thickening fluid is defined as follows.
We imagine an ensemble of mesoscopic elastic elements, each having a local
strain variable $l$. We consider only simple shear strains, and neglect
normal stresses, so that the only nontrivial stress is the corresponding
shear stress $\sigma$. (These assumptions reduce the problem to a scalar one.)
Let us define $P(l,t)$ to be the probability density function
of elements which have a local strain $l$ at time~$t$. We assume $P(l,t)$
evolves in time according to two distinct mechanisms: homogeneous shearing
at a rate~$\gd$, and the yielding of elements at a rate $\Gamma(l)$
per unit time. 
In calculating the stress, the elements are supposed to behave elastically
between yield events, so that the local stress is just~$kl$,
where $k>0$ is an elastic constant. The global stress $\sigma$ is
simply the arithmetic mean of the local stresses, or
$\sigma=\langle kl\rangle=k\langle l\rangle$.
(Here the angled brackets represent an instantaneous
average over $P(l,t)$.) We ignore completely the fact that,
in practice, the strain rate $\gd$ can vary locally in space
in response to each mesoscopic element having a different local stress.

In the present work, $\Gamma(l)$ is the same for all elements,
which simplifies the analysis significantly. A more comprehensive model,
in which elements each have their own `yield strain' $E$ is studied in a
longer paper \cite{Longer}, where further details of some of our
calculations can also be found. Both models are closely related
to the (shear-thinning) SGR model of Sollich {\em et al.}~\cite{Sollich},
but we will choose a very different form of $\Gamma(l)$ from that of SGR.

The master equation for $P(l,t)$ is then
\begin{equation}
\frac{\partial P}{\partial t}+
\gd
\frac{\partial P}{\partial l}=
-\Gamma(l)P
+\hop(t)\,\delta(l)
\label{e:master}
\end{equation}
The second term on the left--hand side
represents the increase in local strains $l$ according to
the spatially uniform strain rate~$\gd$.
The two terms on the right hand side describe the
birth and death of elements with strain~$l$, respectively,
where $\delta(l)$ reflects the assumption that newly--yielded
elements are unstrained.
The total rate of yielding $\hop(t)$ is defined by
\begin{equation}
\hop(t)=
\int_{-\infty}^{\infty}{\rm d}l\,
\Gamma(l)P(l,t)
=
\langle\,\Gamma(l)\,\rangle
\label{e:yield}
\end{equation}

The key remaining ingredient of our model is the hopping rate $\Gamma(l)$.
We parameterise this as follows:
\begin{equation}
\Gamma(l)=
\Gamma_{0}\exp\left[-\left(E-kl^{2}/2\right)/x(l)\right]
\label{e:gamma}
\end{equation}
This is a pseudo-activated form, in which each element is subject to
an effective temperature $x$, and attempts (at a rate $\Gamma_0$) to
hop over a barrier of height $E -\frac{1}{2}kl^{2}$. The latter expression
allows for the lowering of the elastic yield barrier by the imposition of
strain; on its own this will always lead to shear thinning. 

However, the novel feature in the current model is to allow $x(l)$ to be a
decreasing function of $l$. This reflects the intuition that, even if a
small local stress or strain always promotes yield, a large enough one may
jam an element against its neighbours causing the jump rate to fall.
Other than in steady state (see below) we have been unable to find an
analytic solution to (\ref{e:master}) for any non-trivial $x(l)$.
Instead it has been numerically integrated.
With the above as motivation, we chose for numerical purposes relatively
simple (piecewise constant) decreasing functions $x(l)$.
The precise choice of $x(l)$ --- in particular,
whether it is smooth or not --- does not seem to qualitatively
influence the results.

\section{Results}

The steady state solution $P_\infty(l)$ of (\ref{e:master}) is found as
(setting $k=\Gamma_{0}=1$ from now on, for convenience)

\begin{equation}
P_\infty(l) = \omega_\infty \gd^{-1} \exp\left[-\gd^{-1}f(l) \right]
\label{e:st3}
\end{equation}
where $f(l)\equiv\int_0^ldl'\,\Gamma(l')$, and $\omega_\infty$,
the asymptotic jump rate, is fixed by normalization of $P_\infty$.
It is straightforward to show from (\ref{e:st3}) that in the
limit of slow flows $\gd\to0^{+}$, the steady state
stress response $\sigma_\infty = \langle l\rangle_\infty$
(in an obvious notation) is always linear:
$\sigma_{\infty}\sim\gd{\rm e}^{E/x(0)}$.
Hence there is no yield stress for any choice of $x(0)$.
This contrasts with a model having an exponential distribution
of barrier heights $E$ for different elements,
which does show onset of a yield stress, connected with the presence
of a glass transition, as $x$ is reduced \cite{Sollich}.
For monodisperse $E$, as here, there is no such transition. 

We now show that the steady state flow curve, for {\em any}
choice of function $x(l)$, has a monotonically increasing $\sigma(\gd)$. 
First, by differentiating $\langle l\rangle_{\infty}$ and using
(\ref{e:st3}), we find 
\begin{equation}
\frac{\partial\sigma_{\infty}}{\partial\dot{\gamma}}
=
\frac{\sigma_{\infty}}{\hop_{\infty}}
\frac{\partial\hop_{\infty}}{\partial\dot{\gamma}}
-\frac{\sigma_{\infty}}{\dot{\gamma}}
+\frac{1}{\dot{\gamma}^{2}}
\left\langle
lf
\right\rangle_{\infty}
\label{e:mono_local}
\end{equation}
Similarly, the normalisation integral for $P_\infty$ can be
differentiated with respect to $\gd$ to give
\begin{equation}
\frac{1}{\hop_{\infty}}
\frac{\partial\hop_{\infty}}{\partial\dot{\gamma}}=
\frac{1}{\dot{\gamma}}
-\frac{1}{\dot{\gamma}^{2}}
\left\langle
f
\right\rangle_{\infty}
\end{equation}
Combining these two expressions
(with $\sigma_{\infty} = \langle l \rangle_{\infty}$) produces
\begin{equation}
\dot{\gamma}^{2}\,
\frac{\partial\sigma_{\infty}}{\partial\dot{\gamma}}
= \langle lf \rangle_{\infty} - \langle l \rangle_{\infty}
\langle f \rangle_{\infty}
\label{e:mono_ans}
\end{equation}
which is positive since $f(l)$ is monotone increasing.
Thus there are no regions on the flow curve with negative slope,
and for no choice of $x(l)$ 
does one expect any shear-banding instability to arise.
However, the argument does not prove that a steady state is ever
actually achieved.
We have found that it is reached under conditions of
imposed strain $\gamma(t)$, but
{\em not} under an imposed stress $\sigma$ (for appropiate $x(l)$),
for which the model exhibits temporal
oscillations instead.

This numerical finding (detailed below) is also supported by a linear
stability analysis for perturbations about the steady state flow,
described fully in~\cite{Longer}. Such an analysis shows that the
transient behaviour close to steady state is never a real exponential
decay but always has an oscillatory component. When the decay rate of
these oscillations changes sign, the flow becomes unstable to permanent
oscillation.

\section{The oscillatory regime}

The time evolution of the system at fixed stress was studied numerically
by methods described in \cite{Longer}. (The problem is not standard and
significant amounts of computer time were required to achieve reliable
results.) A range of different functions $x(l)$ and of $E$ values were
tried. Varying $\sigma$ over a wide range of values in each case reveals
that steady flow is always reached for sufficiently small and sufficiently
large imposed stresses; but for intermediate $\sigma$,
an oscillatory regime is observed whenever $x(l)$ is strongly
enough decreasing. It is impractical to explore the entire function space
for $x(l)$ so at present the precise criteria for the presence of
oscillations are not known.

\begin{figure}
\centerline{\psfig{file=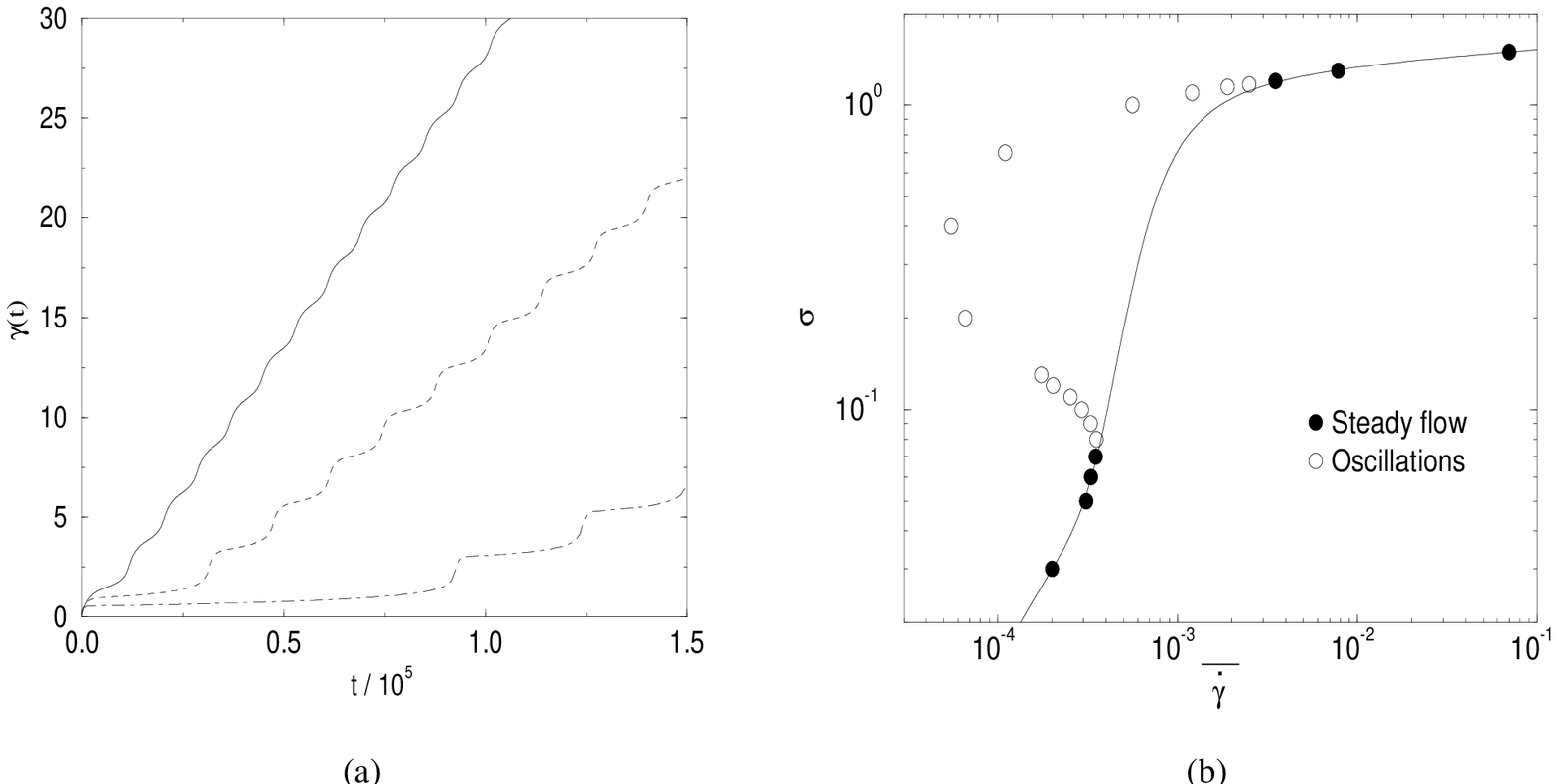,width=15cm}}
\caption{(a)~The strain $\gamma(t)$
under an imposed stress $\sigma=0.1$ (solid line),
0.13 (dashed line) and 0.2 (dot-dashed line)
for $x(l)=1$ for $l<0.4$ and $x=0.4$ for $l>0.4$.
The system was initially unstrained and $E=5$.
(b)~The mean strain rate $\mg$ against the imposed stress $\sigma$
for the same system.
The solid and open circles represent steady and oscillatory
solutions, respectively, as observed from the simulations.
The sizes of the circles are larger than the error bars.
For comparison, the theoretical flow curve for the steady
state solution (\ref{e:st3}) is plotted as a solid line.
}
\label{f:osc_nonmon_eg}
\label{f:osc_flowcurve}
\end{figure}

Within the oscillatory regime, we have so far found only single-period
oscillations with no sign of period doubling or other more complex
nonsteady behaviour that might point to the onset of rheo-chaos in
some parts of the parameter space.  
Some examples of the oscillatory behaviour are given in
Fig.~\ref{f:osc_nonmon_eg}(a).
To obtain these results we took $x=1$ for $l<0.4$ and $x=0.4$ for $l>0.4$;
in addition we set $E=5$. The strain $\gamma(t)$ is shown for
three different
imposed~$\sigma$. On close inspection the oscillations, after a brief
transient, appear perfectly periodic. This applies not only in the stress
but in the underlying distribution $P(l,t)$, which determines the
future evolution. Hence there seems little doubt that these oscillations
will persist indefinitely. 

From Fig.~\ref{f:osc_nonmon_eg}(a) it is remarkable that the {\em mean}
strain rate $\mg$, defined as $\gd$ averaged over a single period
of oscillation, is clearly a {\em decreasing} function of~$\sigma$,
in complete contrast to the monotonic, steady state, flow curve.
The time averaged values $\mg$ as read off from the simulations
are plotted against $\sigma$ in Fig.~\ref{f:osc_flowcurve}(b),
overlayed with the steady state flow curve.
Within the oscillatory regime, the $\mg$
line deviates from the flow curve to the extent that
it loses monotonicity over a large interval of~$\sigma$.
If the oscillations were not detected but only the average
behaviour measured, this would resemble the flow curve of a
discontinuously shear-thickening system close to a jamming transition,
although in that case shear banding could be expected
as well. (Such a flow curve is indeed found in models where $x$
depends not on the local stress $l$ but only on its global mean,
$\sigma$; see \cite{Longer}.)

Upper and lower transition points between oscillatory
and steady flow can be seen in the figure.
At each transition point $\sigma=\sigma_{\rm c}$,
the period of oscillation remains finite
whilst the amplitude vanishes smoothly according to
$|\sigma-\sigma_{\rm c}|^{\alpha}$ with $\alpha>0$.
As we discuss in our longer work~\cite{Longer},
technical difficulties have prevented us from
reliably fixing the asymptotic values of~$\alpha$.
In particular we have been unable to rule out 
$\alpha=0.5$ for both transitions, 
as expected for a Hopf bifurcation~\cite{Glendinning,Jacques}
(although since the control parameter
is an integral constraint,
it is not clear to us if
it should be a Hopf bifurcation).

\begin{figure}
\centerline{\psfig{file=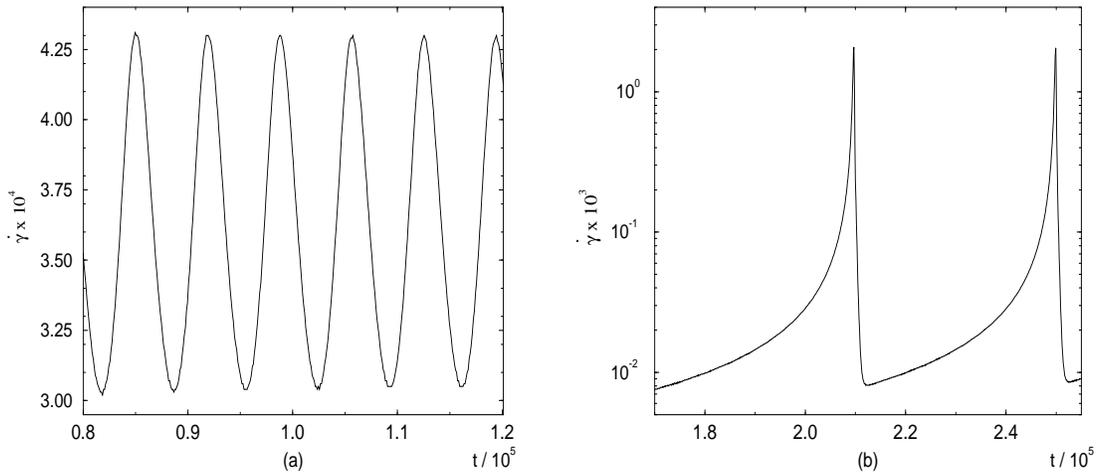,width=15cm}}
\caption{Examples of the oscillatory strain response to a constant
stress for the same $x(l)$ as in Fig.~\ref{f:osc_flowcurve}.
The imposed stresses are
(a)~$\sigma=0.075$, which is just above
the threshold value for steady flow,
and
(b)~$\sigma=0.4$, which
is well into the oscillatory regime.
Note that the axes in (b) are semilogarithmic.
}
\label{f:osc_sine}
\end{figure}

Though always perfectly periodic, the waveform of the oscillations
varies with~$\sigma$. Close to either transition,
the oscillations are near-sinusoidal,
as demonstrated in Fig.~\ref{f:osc_sine}(a).
Further into the oscillatory regime,
$\gd$ can no longer be decomposed into a single harmonic,
but instead approaches a waveform in which most of the
variation in $\gd$ is compressed into
a small fraction of the total period of oscillation.
An example is given in Fig.~\ref{f:osc_sine}(b).
This behaviour at first appears to be similar to the
`stick--slip' motion observed in systems such as
earthquakes~\cite{earthquake},
ultra-thin liquid films~\cite{thinliquids}
and granular media~\cite{granstickslip1,granstickslip2}.
However, the underlying physics in our model
seems to be somewhat different to these examples.
Indeed, stick--slip is usually viewed as
a surface phenomenon,
whereas the model studied in this paper has no surface and
describes only bulk effects.
Irrespective of the waveform, the product
of $\mg$ and the period of oscillation $T$
is approximately constant,
$l^{*}=\mg T\approx2.3$ for this example.
This will be explained below in terms of the
evolution of $P(l,t)$.


\section{Mechanism of oscillation}

\begin{figure}
\centerline{\psfig{file=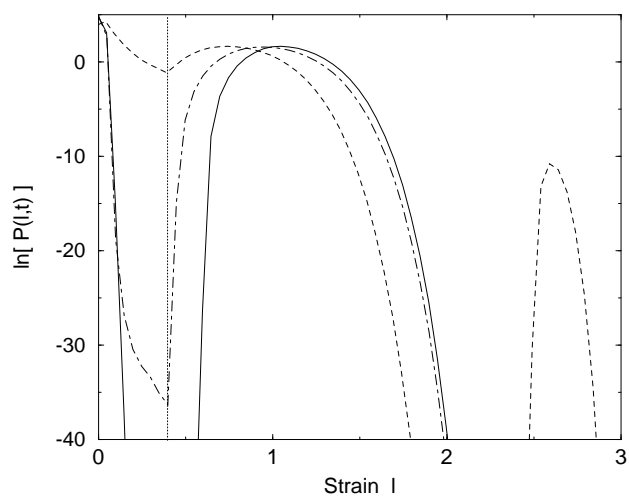,width=9cm,angle=270}}
\caption{Snapshots of $P(l,t)$ in a monodisperse system
with $E_{1}=5$, and $x(l)=1$ for $l<0.4$, 0.4 otherwise.
The imposed stress was $\sigma=0.2$.
The times of each plot are $t=1.1\times10^{5}$ (solid line),
$t=1.25\times10^{5}$ (dashed line) and
$t=1.3\times10^{5}$ (dot-dashed line),
corresponding to before, during and just after
the peak in $\gd$, respectively.
The vertical dotted line is where $x(l)$ changes value.
Note that this corresponds to the lower line in Fig.~\ref{f:osc_nonmon_eg}.
}
\label{f:snapshots_mono}
\end{figure}

Snapshots of $P(l,t)$ during a single period of
oscillation are given in Fig.~\ref{f:snapshots_mono}.
The mechanism behind the oscillations
can be qualitatively described in terms of two coexisting
populations of elements --- a `hot' population of unstrained
or slightly strained elements with $l\approx0$, and hence
a high effective temperature $x(l)$;
and a `cold' population of elements with $l\gg0$
and therefore a low $x(l)$.
Starting at a time when $\gd(t)$ is near to is minimum value,
corresponding to the first snapshot in the figure,
the stress--weighted yield rate $\langle l\Gamma(l)\rangle$ for
both populations of elements is low.
This is because the hot elements have small~$l$,
whereas the cold elements have a low yield rate $\Gamma(l)$.
Thus the strain rate $\gd$ required to maintain the
imposed constant mean stress~$\sigma$,
which is $\gd(t)=\langle l\Gamma(l)\rangle$
(as seen by multiplying (\ref{e:master}) by $l\,{\rm d}l$ and integrating),
is also low.

Although $\gd(t)$ is small in this state, it is nonetheless non--zero,
and therefore the strains $l$ of the cold elements
increase according to $\dot{l}=\gd$.
This decreases their effective energy barrier $E-l^{2}/2$,
increasing their hop rate and therefore
the rate of stress lost due to to cold elements yielding.
Thus $\gd(t)=\langle lP(l)\rangle$ will also increase,
which accelerates the rate at which the cold elements yield,
and so on.
This description is that of a {\em positive feedback loop}
that causes $\gd$ to increase at ever faster rates until
all of the cold elements have been depleted.
The second snapshot in Fig.~\ref{f:snapshots_mono}
show the state of the system shortly before this has happened.

While $\gd(t)$ is high,
the hot elements are becoming rapidly strained and
have little time to yield before crossing over into
the cold region with large~$l$.
Thus $P(l,t)$ is flat for small~$l$.
As $\gd(t)$ again decreases,
this flat part of the distribution will start to decay
as the elements within it yield.
However, the yield rate depends on~$x(l)$,
and since this changes to a lower value at $l=0.4$,
$P(l,t)$ will decay more rapidly for $l$ smaller
than $0.4$ than for $l$ greater than $0.4$.
Thus a dip will occur around the point $l\approx0.4$,
which can clearly be seen in Fig.~\ref{f:snapshots_mono}.
This dip becomes more pronounced with time
until the system can again be viewed as
coexisting `hot' and `cold' populations,
when the cycle begins again.

A possible interpretation of the parameter $l^{*}$,
defined previously, is that
it is the amount by which the system needs to
be strained until the positive feedback loop just described
starts to dominate the system behaviour, causing it to
`reset' to the start of its cycle.
If this is correct,
then $l^{*}$ should correspond to the point at which
highly--strained elements have the same yield rate as
unstrained elements,
{\em i.e.} $\Gamma(0)\approx\Gamma(l^{*})$.
Rough calculations based on this assumption give
$l^{*}\approx\sqrt{2E[1-x(\infty)/x(0)]}$,
which for this example predicts $l^{*}\approx2.4$, in fair agreement
with the observed value.

\section{Discussion}

Within a simple model of shear thickening, we have found regimes where
steady flow at constant imposed stress is unstable and temporal
oscillation occurs instead. Known instances of such rheological
instability have often been
explained in terms of the spatial coexistence of subpopulations, or phases.
For instance, the temporal oscillations in viscosity observed in wormlike
micelles under an imposed stress was attributed to
a slowly fluctuating interface between
a fluid phase and shear-induced structures~\cite{Hu}.
For surfactant solutions in the lyotropic lamellar phase,
it was attributed to coexisting
ordered and disordered phases~\cite{Jacques}.
By contrast, the temporal oscillations observed in our model 
arise even though they are by assumption spatially homogeneous,
and the flow curves are everywhere monotonic. It may be that, in practice,
such rheo-oscillations will always be coupled to spatial heterogeneity of
some sort. Experiments are not advanced enough to indicate whether this is
the case.

One distinguishing feature of our equations is the role of memory 
effects, rather than inertia, in allowing oscillations. In our model,
the memory resides in $P(l,t)$ which can have different shapes for the
same applied stress $\langle l \rangle$. There may thus be a link to
work on `delay differential equations' in which a first order differential
equation for a state variable contains a feedback term that depends on the
same state variable at an earlier time \cite{glassmackey}. Such equations
are capable of showing a range of instabilities, including chaos, despite
the absence of inertia. The nature of this link remains a topic of current
research. Its exploration might help answer the following question:
what features, if any, could be added to our simple rheological model of
shear thickening to allow rheo-chaos to arise {\em without} coupling to
spatial degrees of freedom \cite{Keunings}. This could in turn help answer
another important question: what is the dimension of the strange attractors
that arise in rheo-chaos \cite{Sood2}?
In conclusion
we suggest that the study of temporal (or spatiotemporal) pattern formation
in non-Newtonian flows (at effectively zero Reynolds number) will
form a major topic of experimental and theoretical research in the
coming years.


\noindent{\em Acknowledgements:} The authors would like
to thank Suzanne Fielding for   
stimulating discussions concerning this work. 
AA also wishes to thank the University of Edinburgh
for the hospitality that allowed this work to start.
DAH was funded by EPSRC(UK) grant no. GR/M09674.



\end{document}